\documentclass[preprint,12pt,sort&compress,times]{elsarticle}
\journal{Physics Letters B}
\usepackage[breaklinks,colorlinks=true]{hyperref}
\usepackage{amsmath}
\usepackage{amssymb}
\usepackage{bm}
\usepackage{physics}

\newcommand{\rme}{\mathrm{e}}
\newcommand{\rmd}{\mathrm{d}}
\newcommand{\rmi}{\mathrm{i}}
\newcommand{\Nc}{N_{\text{c}}}

\newcommand{\iomega}{\Omega_{\text{I}}}

\newcommand{\hmu}{\hat{\mu}}
\newcommand{\hnu}{\hat{\nu}}

\begin{document}

\begin{frontmatter}

\title{Imaginary Rotating Gluonic Matter at Strong Coupling}

\author[1]{Kenji Fukushima}
\author[1]{Yusuke Shimada}

\affiliation[1]{organization={Department of Physics, The University of Tokyo},
 addressline={7-3-1 Hongo, Bunkyo-ku},
 city={Tokyo},
 postcode={113-0033},
 country={Japan}}

\begin{abstract}
  We write down an effective theory of the Polyakov loop to investigate the deconfinement phase transition of imaginary-rotating gluonic matter using the strong-coupling expansion.  We find the strength of the nearest-neighbor Polyakov-loop interaction modified by the sum of contributions involving the chair-type loops along the temporal direction.  Our results show that the deconfinement transition temperature increases with increasing imaginary angular velocity, which agrees with the predictions from the models and the high-temperature perturbative calculations.
\end{abstract}

\end{frontmatter}

\section{Introduction}

Color confinement is the most fascinating feature of Quantum Chromodynamics (QCD).
At finite temperature $T$, if the system is pure gluonic matter (i.e., quark masses are infinitely large), the realization of center symmetry distinguishes confinement and deconfinement.
The spontaneous breaking of center symmetry is characterized by the expectation value of a non-local operator called the Polyakov loop~\cite{Polyakov:1975rs}, which was originally addressed in an effective theory of gluons in the strong-coupling limit in Ref.~\cite{Polyakov:1978vu}.  Since these early days~\cite{Wilson:1974sk,Polyakov:1978vu}, the strong-coupling expansion has provided us with a useful theoretical tool to reveal the mysteries of confinement.

The Polyakov loop effective action has also been derived from the opposite limit of deconfinement matter within high-temperature QCD perturbatively~\cite{Weiss:1980rj,Weiss:1981ev,Gross:1980br,KorthalsAltes:1993ca,Gocksch:1993iy}, and center symmetry is indeed broken there by the appearance of the nonzero Debye screening mass (for a review, see, e.g., Ref.~\cite{Fukushima:2017csk}).

It is not only the temperature but also other environmental parameters that allow for accessing the mechanism of deconfinement phase transitions.  Several examples of such extreme probes include the baryon density, the isospin density, the magnetic field, the (imaginary) electric field (see Ref.~\cite{Endrodi:2024cqn} for a comprehensive review), the scalar curvature, and the (imaginary) angular velocity, etc.
In particular, among various probes, the angular velocity $\omega$ has drawn special attention from both experimental and theoretical sides.
The heavy-ion collision experiment has confirmed a nonzero value of the global spin polarization, from which $\omega \sim 10^{22}\ \mathrm{s}^{-1}$ has been deduced~\cite{STAR:2017ckg,Becattini:2013fla}.
Since this confirmation, the phase diagram of fast-rotating matter has become a realistic possibility, and the rotation effects on the QCD phase transitions have been intensively studied.
In the chiral sector, the effective model approach clearly demonstrates an analogy between the rotation effect and the finite baryon density effect~\cite{Chen:2015hfc}.  Then, the phase diagram of rotating QCD matter has been revealed, showing the decreasing behavior of the chiral transition temperature with increasing density~\cite{Jiang:2016wvv}.  All chiral models even with improved approximations make similar predictions~\cite{Chernodub:2016kxh,Wang:2018sur,Chen:2023cjt,Sun:2023kuu}, except for a fit with $\omega$-dependent coupling~\cite{Nunes:2024hzy}.
The confinement sector is more subtle.  Under some assumptions for the transition criterion, the hadron resonance gas model has concluded that the deconfinement phase transition (or the chemical freezeout line) should be lowered in temperature at larger $\omega$~\cite{Fujimoto:2021xix}.  In the holographic QCD model, the deconfinement transition is nothing but the geometrical phase transition to the thermal AdS black hole, and the qualitative features are consistent with other model predictions~\cite{Chen:2020ath,Braga:2022yfe,Yadav:2022qcl,Wang:2024szr,Zhao:2022uxc}.

The theoretical advantage in considering the rotation is that, although the rotation effect has similarities with the finite density effect, the rotation is coupled with not only quarks but also gluons.  Therefore, the pure gluonic system in which the deconfinement phase transition can be defined rigorously may also exhibit nontrivial rotation effects.  In our previous studies~\cite{Chen:2022smf,Chen:2024tkr}, we analyzed the rotation effect in the pure gluonic system at high temperatures enough to justify the perturbative calculation.  This strategy should lead to the first-principles answer in the pure gluonic sector of QCD, and the results are again consistent with other model predictions.
It is worth mentioning that a novel mechanism of confinement has been found from the perturbative calculation with imaginary-valued angular velocity, i.e., $\iomega=-\rmi \omega$.

In contrast, the lattice-QCD numerical simulations of rotating QCD matter have reported counter-intuitive results.  The first lattice results were reported in a seminal work by Yamamoto and Hirono~\cite{Yamamoto:2013zwa} in which the orbital and the spin angular momenta of gluons and quarks were measured by the analytical continuation technique from the imaginary rotation, $\iomega=-\rmi\omega$, to evade the sign problem.
Then, the phase diagram research has been pioneered in Ref.~\cite{Braguta:2021jgn}, followed by more detailed studies~\cite{Chernodub:2022veq,Braguta:2023iyx,Yang:2023vsw}.  These lattice studies have concluded that the critical temperature should be raised by the real rotation effects, which contradicts the model predictions.  Also, we would emphasize that the lattice results qualitatively disagree with the perturbative QCD calculation, which must be a valid approximation if the temperature is high enough.  There are many discussions about these discrepancies --- subtle relationships between real and imaginary rotations~\cite{Chen:2023cjt,Yang:2023vsw,Cao:2023olg,Jiang:2023zzu,Chernodub:2022veq,Braguta:2023iyx}, hybrid models in support of the lattice results~\cite{Cao:2023olg, Mameda:2023sst, Sun:2024anu, Chen:2024jet,Chen:2024edy}, implementation of non-monotonic critical temperatures~\cite{Jiang:2023zzu, Gaspar:2023nqk}, and so on.

This paper aims to add another QCD-based analysis of rotating gluonic matter.  As mentioned in the beginning, the strong-coupling expansion has played an important role in understanding the QCD phase transitions qualitatively and even semi-quantitatively~\cite{Fukushima:2003vi}.  The technique has been applied to the QCD phase diagram research; see, e.g., a classic formulation~\cite{Green:1983sd} based on the hopping parameter expansion, a first strong-coupling phase diagram in Fig.~4 of Ref.~\cite{Fukushima:2003vi}, an extensive analysis of the phase diagram with parameter dependence~\cite{Kawamoto:2005mq}, and also a strong-coupling Monte-Carlo method in the continuum limit~\cite{deForcrand:2014tha}.
Just recently, we noticed that a study based on a similar idea of the strong-coupling expansion appeared~\cite{Wang:2025mmv}.
While in Ref.~\cite{Wang:2025mmv} they treated the real rotation effect up to the $\omega^2$ order, the present work will deal with the imaginary-rotating effective theory at finite temperature in terms of the Polyakov loop including higher-order terms in $\iomega$ in a systematic expansion of the strong coupling.  We will finally make a quantitative estimate of how much the critical temperature is shifted by the rotation effects.

\section{Polyakov Loop Effective Theory at Strong Coupling}

Here, as a starting point of the strong-coupling expansion on the lattice, we adopt the lattice action proposed in Ref.~\cite{Yamamoto:2013zwa}.  In the Cartesian coordinates with rotation along the $z$-axis, the metric takes the following form,
\begin{align}
    g_{\mu\nu} &= 
    \begin{pmatrix}
        1 & 0 & 0 & y\iomega \\
        0 & 1 & 0 & -x\iomega \\
        0 & 0 & 1 & 0 \\
        y\iomega & -x\iomega & 0 & 1+r^2\iomega \\
    \end{pmatrix} \,, 
\end{align}
where $r = \sqrt{x^2+y^2}$.  The lattice action of gluonic matter at finite imaginary angular velocity, $\iomega$, is
\begin{align}
    S_\mathrm{G} = \frac{1}{g_{\mathrm{YM}}^2}  & \int \rmd^4 x\,
    \tr \Bigl[ (1+r^2\iomega^2)F_{xy}F_{xy}
    + (1+y^2\iomega^2)F_{xz}F_{xz}
    + (1+x^2\iomega^2)F_{yz}F_{yz} \notag\\
    & + F_{x\tau}F_{x\tau} 
    + F_{y\tau}F_{y\tau} 
    + F_{z\tau}F_{z\tau}
    + 2y\iomega F_{xy} F_{y\tau}
    - 2x\iomega F_{yx} F_{x\tau} \notag\\
    & + 2y\iomega F_{xz} F_{z\tau}
    - 2x\iomega F_{yz} F_{z\tau}
    + 2xy\iomega^2 F_{xz}F_{zy}
    \Bigr]
\end{align}
in the continuum theory.
The first six terms have the same tensor structures as those in the action without rotation.  There are additional terms induced by $\iomega$ only in the magnetic terms, while the electric terms are intact.  Usually, these terms are replaced by the plaquettes in the lattice formulation. Since the coefficients depend on space, however, it is convenient to introduce the spatially averaged plaquette as
\begin{align}
    U_{\mu\nu}(n) &= \left( \ \parbox[c]{47pt}{\includegraphics[scale=0.35]{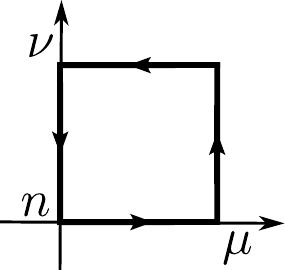}} \right) = U_\mu(n) U_\nu(n+\hmu) U_\mu^{\dag}(n+\hnu) U_\nu^{\dag}(n) \,, \\
    \bar{U}_{\mu\nu}(n) &= \frac{1}{4} \left[ U_{\mu\nu}(n) + U_{\mu\nu}(n-\hmu) + U_{\mu\nu}(n-\hnu) + U_{\mu\nu}(n-\hmu-\hnu) \right] \,.
    \label{eq:U}
\end{align}
Here, $n$ denotes the four-dimensional lattice coordinates: $n = (x,y,z,\tau)$.

The last five terms have combinations of the field strength tensors in an unconventional way to form the chair-type loops~\cite{Iwasaki:1983iya}.
For example, to generate a term, $2F_{xy}F_{yz}$, we should expand loops of $U_{xy}$ and $U_{yz}$, but this procedure also generates unwanted terms such as $F_{xy}^2 + F_{yz}^2$.
Then, it is essential to define the averaging combination as
\begin{align}
    {V}_{\mu\nu\rho}(n) &= \frac{1}{4} 
    \left( \ \parbox[c]{44pt}{\includegraphics[scale=0.3]{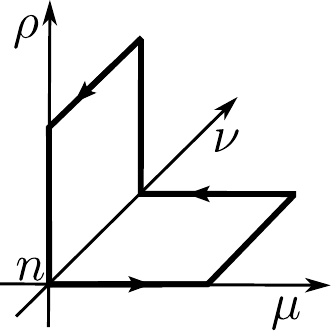}}\ 
    - \ \parbox[c]{44pt}{\includegraphics[scale=0.3]{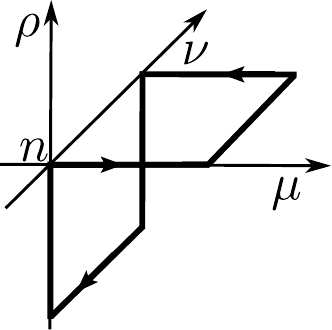}}\ 
    + \ \parbox[c]{44pt}{\includegraphics[scale=0.3]{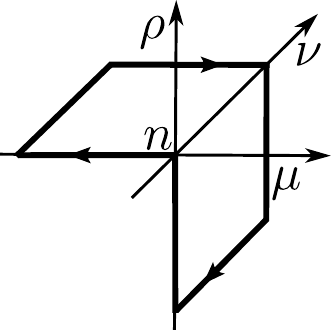}}\ 
    - \ \parbox[c]{44pt}{\includegraphics[scale=0.3]{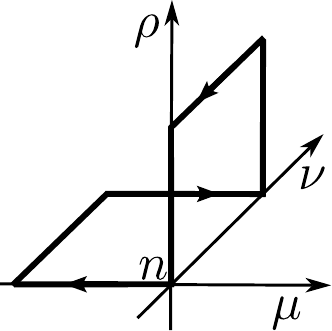}}\ \right)\,, \\
    \bar{V}_{\mu\nu\rho}(n) &= \frac{1}{2} \left[{V}_{\mu\nu\rho}(n) + {V}_{\mu\nu\rho}(n-\hat{\nu})\right] \,,
    \label{eq:V}
\end{align}
where the explicit forms of $V_{\mu\nu\rho}(n)$ should be understood in the same way as in Eq.~\eqref{eq:U}, and they make the spatial average over four different orientations of chair-type loops to eliminate $F_{\mu\nu}^2+F_{\nu\rho}^2$.  Furthermore, it is natural to define $\bar{V}_{\mu\nu\rho}$ to take the average along the remaining $\nu$ direction.
Using these lattice variables, we can express the lattice action with imaginary rotation as
\begin{align}
    S_\mathrm{G} = &\sum_{n} \sum_{(\mu, \nu)} \beta \biggl[
    (1+r^2\iomega^2) \qty(1-\frac{1}{\Nc}\Re\tr \bar{U}_{xy})
    + (1+y^2\iomega^2) \qty(1-\frac{1}{\Nc}\Re\tr \bar{U}_{xz}) \notag\\
    & + (1+x^2\iomega^2) \qty(1-\frac{1}{\Nc}\Re\tr \bar{U}_{yz})
    - \frac{1}{\Nc}\Re\tr (\bar{U}_{x\tau} + \bar{U}_{y\tau} + \bar{U}_{z\tau}) \notag\\
    & - \frac{1}{\Nc}\Re\tr \Bigl( y\iomega \bar{V}_{xy\tau} -x\iomega \bar{V}_{yx\tau}
    + y\iomega \bar{V}_{xz\tau} -x\iomega \bar{V}_{yz\tau} + xy\iomega^2 \bar{V}_{xzy} \
    \Bigr) \biggr]\,,
    \label{eq:lattice_action}
\end{align}
where the overall coefficient is the inverse lattice coupling, i.e., $\beta = 2\Nc/g_{\mathrm{YM}}^2 \ll 1$.

In the strong-coupling expansion, one should expand $\rme^{-S_\mathrm{G}}$ in the power series of $\beta$ and perform the group integration to keep only the color-singlet components.  At finite temperature, one should take account of the Polyakov loop defined by
\begin{equation}
    L(\bm{n}) = \prod_{\tau=0}^{N_\tau-1} U_\tau (\bm{n},\tau)\,,
\end{equation}
which can have a finite expectation value.  Therefore, one should leave $U_\tau(\bm{n},\tau)$ unintegrated so that one can find the effective theory in terms of the Polyakov loop in the sense of the constrained effective action.

\begin{figure}
    \centering
    \includegraphics[width=0.4\columnwidth]{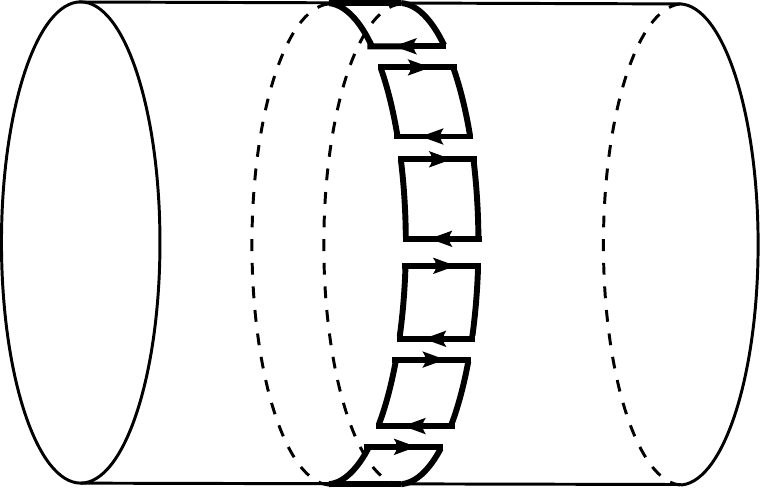}
    \caption{Leading-order contribution to the partition function that results in the effective action of the nearest-neighbor interacting Polyakov loops. The cylinder represents the temporal $S^1$ and the spatial $R^3$.}
    \label{fig:SCEleading}
\end{figure}

In flat spacetime without rotation, the leading-order contributions to the partition function are the configurations of the winding plaquettes along the $\tau$-direction as sketched in Fig.~\ref{fig:SCEleading}; see Ref.~\cite{Green:1983sd}.  Therefore, one obtains~\cite{Fukushima:2002ew,Fukushima:2003fm}
\begin{align}
    S_\mathrm{eff}[U_\tau]
    &= -\ln Z_{\mathrm{eff}}[U_\tau]
    = -\ln\int \mathcal{D}U_i\, \rme^{-S_\mathrm{G}\qty[U_i, U_\tau]} \notag\\
    &= -\ln\Biggl\{ 1 + \sum_{\bm{n}, i}  \biggl(\frac{\beta}{2\Nc^2}\biggr)^{N_\tau} 
    \left[ \tr L^\dag(\bm{n}) \tr L(\bm{n}+\hat{i}) \right] + \cdots\Biggr\} \notag\\
    &\simeq -\sum_{\bm{n},i} J\,\tr L^\dag(\bm{n}) \tr L(\bm{n}+\hat{i}) \,.
    \label{eq:partfunc}
\end{align}
Here, the ellipsis $(\dots)$ represents the terms not involving the Polyakov loop and/or of higher orders in the $\beta$ expansion.
This has an interesting resemblance to the matrix model of the Polyakov loop~\cite{Dumitru:2003hp}.  The nearest-neighbor interaction strength, $J$, is related to the string tension through
\begin{equation}
    J = \rme^{-\sigma a/T}\,,
    \label{eq:JT}
\end{equation}
where $\sigma$ is the string tension and $a$ is the lattice spacing.  This relation is derived from the correlation function of the Polyakov loops and the interquark linear-rising potential.
This effective action represents a simple yet useful model to understand the semi-quantitative properties of the deconfinement phase transition.

In the simplest mean-field approximation, the Polyakov loop has a homogeneous expectation value.  The group integration nature is reflected in the Haar measure, that is, the Vandermonde determinant of the $SU(\Nc)$ group~\cite{Reinhardt:1996fs}.  Then, the mean-field effective potential reads:
\begin{equation}
    V_\mathrm{eff}[L]/T = -6 J |\tr L|^2 - \ln M_\mathrm{Haar}[L]\,,
    \label{eq:Veff}
\end{equation}
where the explicit expression for the Haar measure potential is
\begin{equation}
    M_\mathrm{Haar}[L] = \begin{cases}
        4 - |\tr L|^2 & \text{for $SU(2)$}\,,\\
        27 - 18|\tr L|^2 + 8\Re(\tr L)^3 - |\tr L|^4 & \text{for $SU(3)$}\,.
    \end{cases}
\end{equation}

\begin{figure}
    \centering
    \includegraphics[width=0.7\columnwidth]{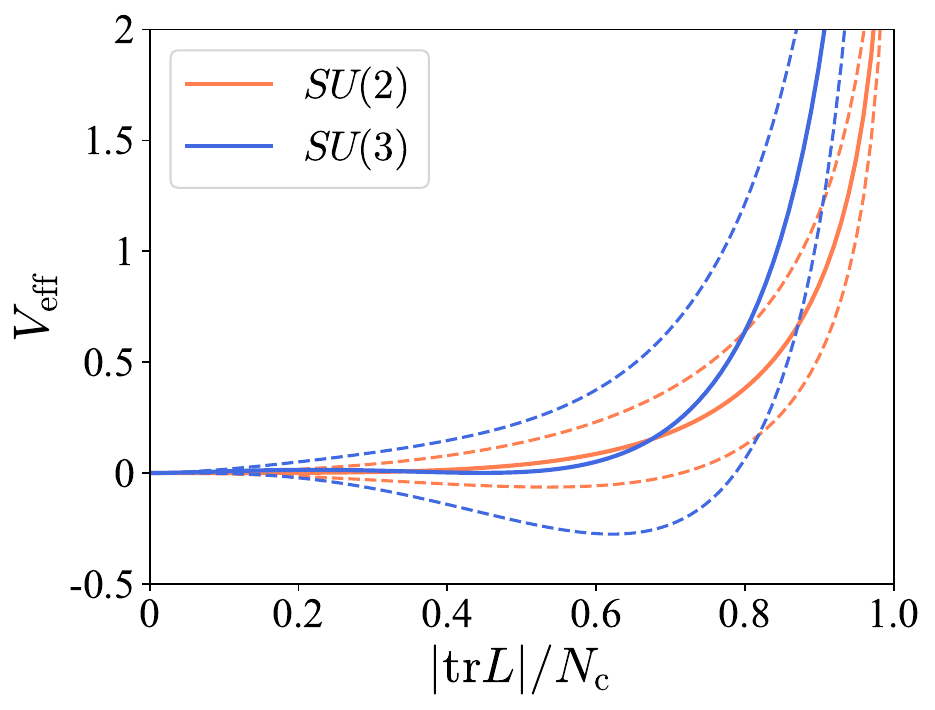}
    \caption{Effective potential shapes at $J=J_\mathrm{c}$ (solid curve) and $J=J_\mathrm{c}\pm 0.1$ (dashed curves) for the $SU(2)$ and the $SU(3)$ cases.  The potential offset is chosen to adjust $V_\mathrm{eff}=0$ at $|\tr L|=0$.}
    \label{fig:transition}
\end{figure}

We note that in both cases of $SU(2)$ and $SU(3)$, in the deconfined phase where $\Nc^{-1}|\tr L|=1$, the Haar measure is vanishing, i.e., $M_\mathrm{Haar}=0$ and thus $\ln M_\mathrm{Haar}$ diverges there.  This means that the perturbative vacuum is unstable in the strong-coupling regime and the confined phase is realized by the Haar measure.  The critical value of $J$ is found to be
\begin{equation}
    6J_\mathrm{c} = \begin{cases}
     0.25 & \text{for $SU(2)$}\,, \\
     0.515 & \text{for $SU(3)$}\,.
    \end{cases}
    \label{eq:Jc}
\end{equation}
For the $SU(2)$ case, the deconfinement phase transition is of second order and the critical condition of vanishing quadratic coefficient immediately leads to $6J_\mathrm{c}=1/4$.  Since the first-order phase transition occurs for the $SU(3)$ case, the critical coupling is an approximate value found numerically.  The potential shapes around $J_\mathrm{c}$ are shown in Fig.~\ref{fig:transition}.

\begin{figure}
    \centering
    \includegraphics[width=0.4\columnwidth]{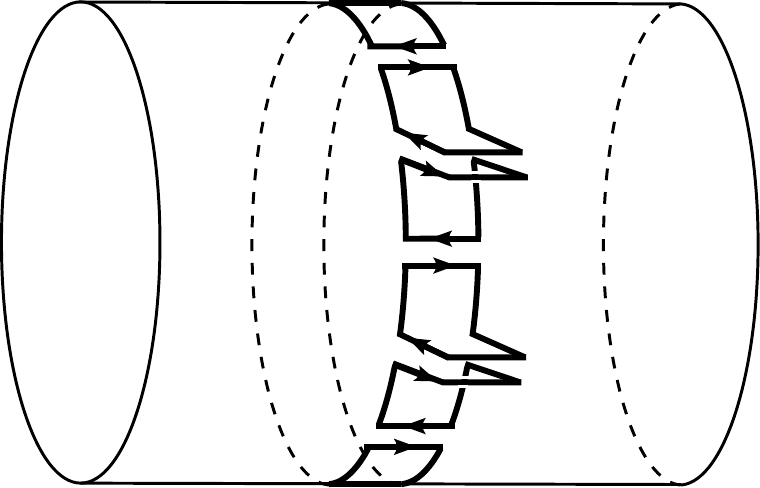}
    \hspace{1em}
    \includegraphics[width=0.4\columnwidth]{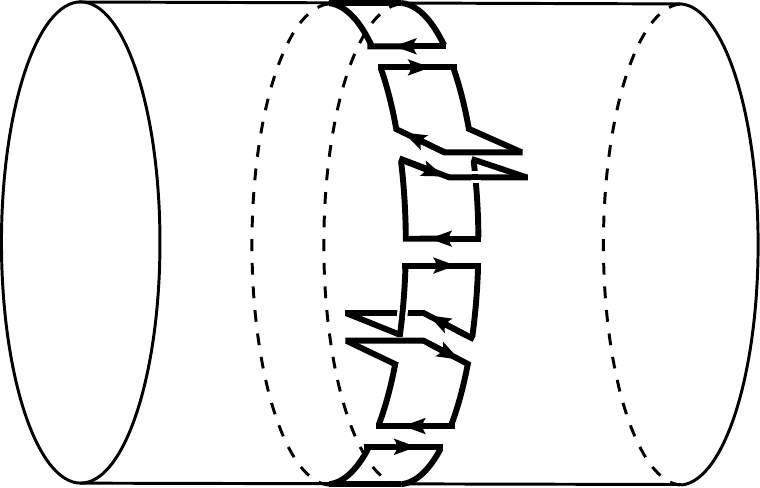}
    \caption{Examples of $\iomega^4$-contributions to the effective action of the same order of $\beta$.  Chair-type loops extend two spatial directions which are schematically represented by the directions parallel and normal to the cylinder surface.  The pairs of chair-type loops can have different orientations as contrasted in the left and the right panels.  The cylinder represents the temporal $S^1$ and the spatial $R^3$.
    }
    \label{fig:SCEleading_rot}
\end{figure}

Our strategy to consider the rotation effects is based on the above-mentioned calculation including $\iomega$-dependent chair-type loops in Eq.~\eqref{eq:V}. One can make a pair of adjacent chair-type loops to make the singlets except for $U_\tau$, and the left panel of Fig.~\ref{fig:SCEleading_rot} is an example of two insertions of such pairs.  Schematically, a spatial direction orthogonal to the Polyakov loop correlation is represented by the direction normal to the cylinder surface.
In this construction, any successive two plaquettes can be replaced with a pair of chair-type loops.  We note that there are two distinct orientations: space-space plaquettes (normal to the cylinder surface) can be directed toward the front or the back as shown in the right panel of Fig.~\ref{fig:SCEleading_rot}.

It is very important to notice that these contributions with multiple pairs of chair-type loops are of the same order in the small $\beta$ expansion.  Therefore, one must take the sum of all the possible insertions of the chair-type loops.
If the number of the inserted pairs of chair-type loops is $k$, there are $N_\tau - 2k$ unchanged plaquettes unless a pair is formed at the periodic edges at $\tau=0$ and $\tau=(N_\tau-1)a$ (where $a$ is the lattice spacing).  Then, the combinatorial factor associated with the insertion is obtained from $\binom{N_\tau-2k+k}{k}$ where $\binom{n}{m}$ is the binomial coefficient.  If one pair is located at the edges of the periodic boundary, there are still $(k-1)$ pairs of the chair-type loops, leaving $(N_\tau-2) - 2(k-1)=N_\tau - 2k$ plaquettes unchanged, whose combinatorial factor is $\binom{N_\tau - 2k + (k-1)}{k-1}=k/(N_\tau-k)\binom{N_\tau-k}{k}$.

\begin{figure}
    \centering
    \includegraphics[width=0.7\columnwidth]{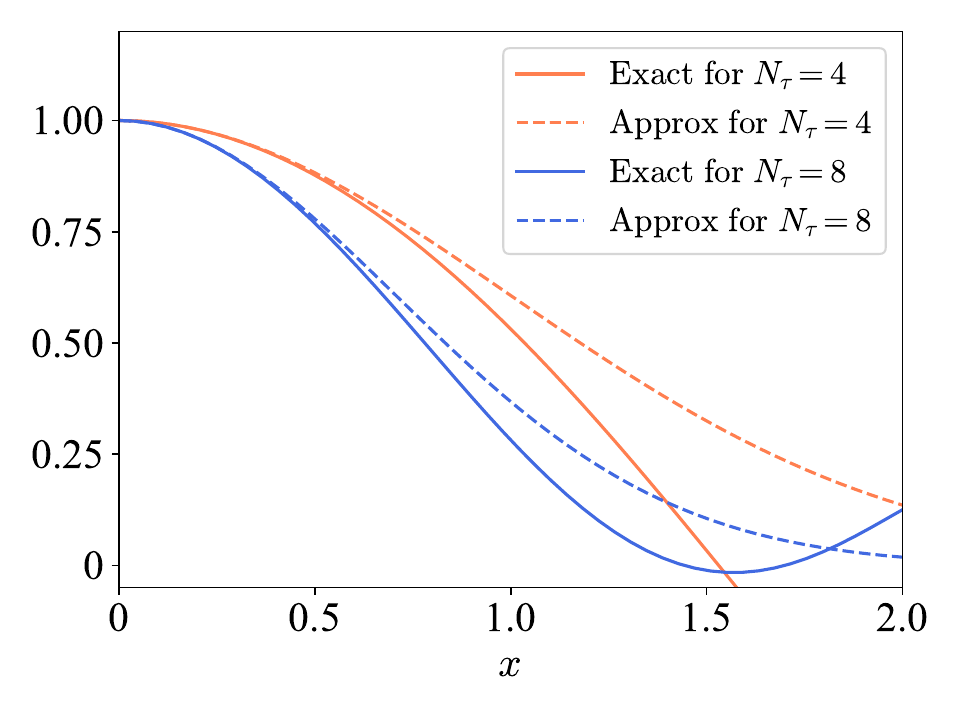}
    \caption{Comparison of the exact sum defined in Eq.~\eqref{eq:funcdef} and an approximation by $\exp\qty(-N_\tau x^2/8)$ for $N_\tau=4$ and $N_\tau=8$.
    }
    \label{fig:approxfunc}
\end{figure}

One can sum all these contributions up, and write down the effective theory of the Polyakov loop as
\begin{align}
    S_\mathrm{eff}[U_\tau] &= -\ln Z_\mathrm{eff}[U_\tau] \notag\\
    &\simeq -\sum_{\bm n} J \biggl[ \tr L^\dag({\bm n}) \tr L({\bm n}+\hat{x}) f(N_\tau, y\iomega) \notag \\
    & \qquad\qquad+ \tr L^\dag({\bm n}) \tr L({\bm n}+\hat{y}) f(N_\tau, x\iomega) \notag \\
    & \qquad\qquad+ \tr L^\dag({\bm n}) \tr L({\bm n}+\hat{z}) f(N_\tau, r\iomega) + (h.c.) + \cdots \biggr] \,,
    \label{eq:partfunc}
\end{align}
where we defined the following function:
\begin{equation}
    f(N_\tau, x) = \sum_{k=0}^{N_\tau/2} \frac{N_\tau}{N_\tau - k} \binom{N_\tau - k}{k}\qty(-\frac{x^2}{8})^k \,.
    \label{eq:funcdef}
\end{equation}
When $N_\tau$ is large or $x$ is small enough, this function is well approximated by an exponential form, $\exp(-N_\tau x^2/8)$, as seen in Fig.~\ref{fig:approxfunc}. 
It is worth noting that none of these functions, $f(N_\tau,x\iomega)$, $f(N_\tau,y\iomega)$, $f(N_\tau,r\iomega)$, is periodic with respect to $\iomega$. Since the loss of the periodicity follows directly from the lattice action, it seems that the lattice action~\eqref{eq:lattice_action} needs some improvements to recover the properties the imaginary-rotating gluonic system should have.
Furthermore, Eq.~\eqref{eq:partfunc} implies that the rotating system has inhomogeneity in space; the results depend on $x$, $y$, and $r$.  In principle, the system should be axially symmetric around the $z$ axis, but the coefficients in Eq.~\eqref{eq:partfunc} are dependent on the azimuthal angle between $x$ and $y$.  This lattice system has only discrete symmetry of $\pi/2$ rotation, reflecting the symmetry kept by the square lattice. For example, the Polyakov-loop interaction coefficients at $\bm{n} = (r/\sqrt{2}, r/\sqrt{2}, 0)$ and $\bm{n} = (r, 0, 0)$ are different, though they should be identical due to axial symmetry.  This behavior is attributed to the lattice discretization artifact that the square lattice is not compatible with the rotation symmetry.
In the continuum limit, this artifact could be resolved by exponentially small coefficients that should be path-integrated over the distance between two separated Polyakov loops.
In other words, the problem of one-lattice spacing has the largest anisotropy.
Here, to suppress the lattice artifact, we shall choose the symmetric point at $\bm{n} = (r/\sqrt{2}, r/\sqrt{2}, 0)$ in what follows below.

\section{Imaginary Rotation Dependence in the Critical Temperature}

Now that we have found the effective theory including imaginary rotation, let us estimate the change in the critical temperature.  To go to realistic quantitative details, we will work in the $SU(3)$ case here.

For rotating systems, one should note that the Polyakov loop interaction over one-lattice spacing is anisotropic except for discrete symmetry under the exchange of $x$ and $y$.  It is a natural expectation that the square-lattice artifact is minimized at the symmetric point of $x=y=r/\sqrt{2}$.  Then, although the Polyakov loop expectation could be a function of $x$ and $y$, we can approximately set $L(\bm{n})\simeq L(\bm{n}+\hat{x}) = L(\bm{n}+\hat{y})$.  Then, the mean-field effective potential in Eq.~\eqref{eq:Veff} is modified as $J\to \tilde{J}(N_\tau,\iomega)$ with
\begin{equation}
     \tilde{J}(N_\tau, \iomega) = \frac{1}{3} \left[ f\qty(N_\tau, \frac{r\iomega}{\sqrt{2}}) + f\qty(N_\tau, \frac{r\iomega}{\sqrt{2}}) + f(N_\tau, r\iomega) \right] J\,,
\end{equation}
as a result of the average of the coefficients in Eq.~\eqref{eq:partfunc}.

\begin{figure}
    \centering
    \includegraphics[width=0.7\columnwidth]{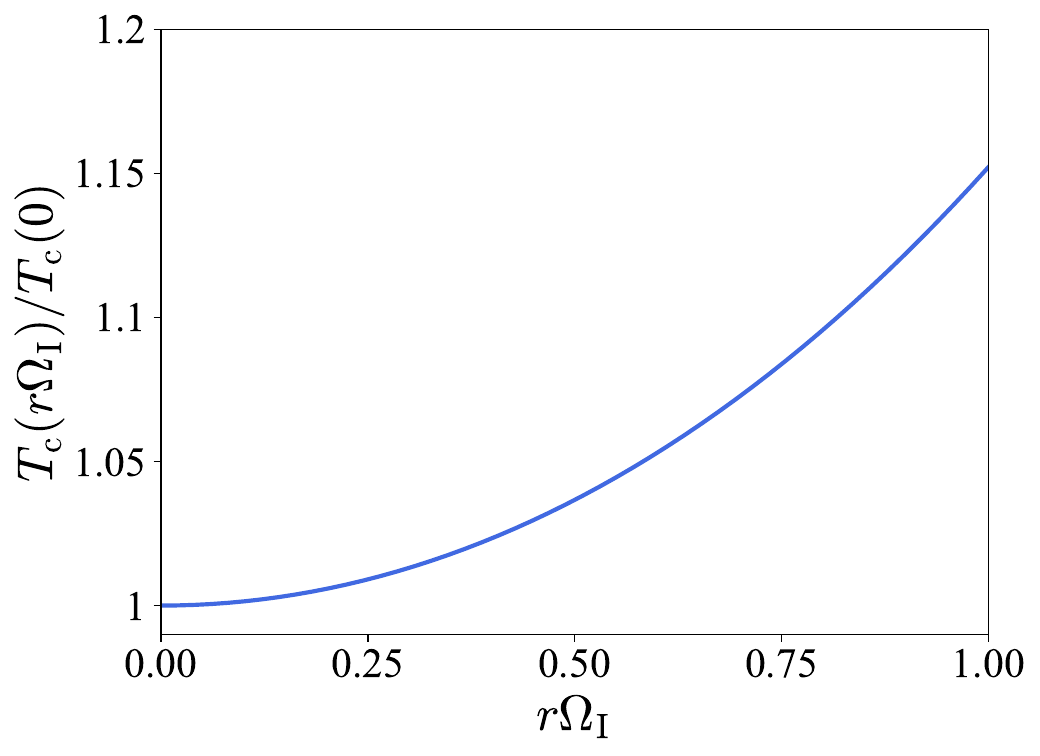}
    \caption{Critical temperature ratio of the deconfinement phase transition as a function of the imaginary velocity, i.e., the distance $r$ times the imaginary angular velocity $\iomega$.
    }
    \label{fig:tempdev}
\end{figure}

Let us recall the critical value of $J$ for the $SU(3)$ case in Eq.~\eqref{eq:Jc}, from which we can immediately get the transition condition,
\begin{equation}
    6\tilde{J}_\mathrm{c}(N_\tau,\iomega) = 0.515\,.
\end{equation}
Because $f(N_\tau,x)$ is a decreasing function as shown in Fig.~\ref{fig:approxfunc}, we understand that $J_\mathrm{c}(\iomega) > J_\mathrm{c}(0)$.  Recalling the relation to the temperature $T$ in Eq.~\eqref{eq:JT}, we conclude that $T_\mathrm{c}$ also increases by the effect of finite $\iomega$.  It is quite intriguing that, using an approximation form of $f(N_\tau,x)\simeq \exp(-N_\tau x^2/8)$ and the relation $J=\rme^{-\sigma a/T}$, we can interpret the imaginary rotation effect as the increasing string tension.  Since the effective string tension gets larger, the critical temperature increases accordingly.

Now, let us further quantify how much the critical temperature is affected by the imaginary rotation.
We adopt the approximation, $f(N_\tau,x)\simeq \exp(-N_\tau x^2/8)$, so our numerical estimates are not very precise for $r\iomega \sim 1$.  In any case, our estimates are only semi-quantitative, and this approximated treatment should suffice for our present purpose.  To express the results in a familiar unit, we fix the strong-coupling parameters from a phenomenological model~\cite{Fukushima:2002ew,Fukushima:2003fm,Fukushima:2003vi}, i.e., $a^{-1}\simeq 433\,\text{MeV}$ and $\sigma\simeq (425\,\text{MeV})^2$, to reproduce the meson mass~\cite{Kawamoto:1981hw}, which leads to a bit small critical temperature, $T_\mathrm{c}(0)=170\,\text{MeV}$ corresponding to $J_\mathrm{c}(0)=0.515$.  In a more sophisticated mean-field analysis, $T_\mathrm{c}$ is found to be substantially larger~\cite{Fukushima:2003fm}.
Then, instead of solving the implicit equation, we use an iterative approximation to obtain our central result of the following expression:
\begin{equation}
    \frac{T_\mathrm{c}(r\iomega)}{T_\mathrm{c}(0)} \simeq \frac{1+r^2\iomega^2/(16\sigma a^2)}{\ln[ J_\mathrm{c}^{-1}(0)(2+\rme^{-r^2\iomega^2/(16T_\mathrm{c} a)})/3] / \ln[J_\mathrm{c}^{-1}(0)]}\,,
    \label{eq:Tcratio}
\end{equation}
where $T_\mathrm{c}$ in the right-hand side is approximated by $T_\mathrm{c}(0)$. 
The increasing behavior of this critical temperature ratio is depicted as a function of $r\iomega$ in Fig.~\ref{fig:tempdev}.  In our choice of parameters, the critical temperature increases by $< 15\%$ as a function of $r\iomega$.
Quantitative details may change with different parameters; if smaller lattice spacing and larger $N_\tau$ are used, the critical temperature ratio could have been more prominently enhanced.
The robust conclusion from our strong-coupling analysis is that the critical temperature changes in a consistent way with the predictions from the effective models and the high-$T$ perturbative calculation.  We also emphasize that our conclusion agrees with the recent strong-coupling study in Ref.~\cite{Wang:2025mmv} at the qualitative level.

The conflict against the lattice numerical simulation is unexpectedly striking.  Usually, the strong-coupling approach works nicely (sometimes even too nicely) to capture the qualitative features of non-perturbative QCD observables.  Our qualitative conclusion about the effects of the imaginary rotation is applied to not only $SU(3)$ but other gauge groups.  Then, Eq.~\eqref{eq:Tcratio} should hold for other gauge groups with some different numbers determined by the different Haar measure.

One can also regard the results in Fig.~\ref{fig:tempdev} as the $r$ dependence for a fixed $\iomega$.  Then, our results are qualitatively consistent with what may result from the Tolman-Ehrenfest effect.  If the critical line is fitted in polynomial form in terms of $\iomega^2$, the analytical continuation with $\iomega^2\to -\omega^2$ would lead to a decrease in the critical temperature with increasing $r\omega$.  The velocity becomes larger for larger $r$ from the rotation center, and then the rotating fluid element is more Lorentz contracted.  Accordingly, the energy density, which is not a Lorentz invariant, increases.  Therefore, the rotation affects the local temperature, and the critical temperature should decrease.
This is again inconsistent with the lattice numerical calculations.

One may wonder that Eq.~\eqref{eq:Tcratio} may underestimate the spin helicity effects $r=0$, which is seemingly missing in our formula.  However, the adjacent interaction of the Polyakov loop originates from the spin-1 nature of the gauge fields, and it is a subtle problem how to place such adjacent objects at the rotation center.  In the vicinity of the rotation center, the square-lattice artifact is severe, and we should probably employ an improved lattice action to recover axial symmetry.

\section{Conclusions}
We analyzed the properties of imaginary-rotating gluonic matter using the strong-coupling expansion on the lattice in which the rotation effect is introduced by the metric into the lattice action.

We calculated the partition function leaving the Polyakov loop unintegrated and explicitly wrote down an effective theory of the Polyakov loop in the form of the nearest-neighbor interaction system.  In our treatment of the systematic expansion of the small inverse coupling, we took account of all higher-order terms with respect to imaginary angular velocity $\iomega$.  In fact, we should make the resummation of all possible insertions of chair-type loops along the temporal $\tau$-direction.  Interestingly, when the lattice size in the $\tau$-direction is sufficiently large or $r\iomega \ll 1$ (where $r$ is the distance from the rotation center), we numerically found that the final sum of such contributions can be well approximated by a simple exponential function of $r\iomega$, which allows us to interpret the imaginary-rotation effect as the increasing modification in the string tension.
We reported the ratio of the critical temperature to show the dependence on the imaginary angular velocity.  In conclusion, the ratio increases as $r\iomega$ increases, which supports the predictions from effective models as well as those from the high-temperature perturbative calculation.

It should be noted that our final results do not maintain the periodicity, $\iomega/T \sim \iomega/T + 2\pi$, as is expected for imaginary-rotating field theories in general~\cite{Chernodub:2020qah,Chen:2022smf}.  This periodicity is a robust consequence of the thermodynamic property as follows:  Rotating systems induce an energy shift by $\bm{\omega}\cdot\bm{\hat{J}}$ with the total angular momentum operator $\bm{\hat{J}}$.  This is generic simply from the density operator.  Therefore, the topological operator, $\exp(\bm{\omega}\cdot\bm{\hat{J}}/T)$, should be inserted into the partition function.  For imaginary angular velocity, $\bm{\omega} = i\bm{\iomega}$, this operator generates a rotation, and $\iomega/T$ is an angular variable.  Since the angular variable is $2\pi$ periodic, the systems must be periodic with respect to $\iomega/T \in \mathbb{R}/2\pi\mathbb{Z}$.  The lack of the expected periodicity seems to be attributed to the lattice action.  As pointed out in Ref.~\cite{Chen:2015hfc}, the angular velocity is similar to the chemical potential, and if this analogy is to be applied to the lattice formulation, the rotation angular velocity may be introduced in a gauge-invariant way as carefully addressed in the lattice formulation; see Ref.~\cite{Hasenfratz:1983ba}.

In addition to the periodicity, the square-lattice formulation also breaks rotational symmetry except for discrete $\pi/2$ rotation.  In our calculations, the effective potential has the modification factors such as $f(N_\tau, r\iomega)$, $f(N_\tau, x\iomega)$, and $f(N_\tau,y\iomega)$, which depend on not only the distance but also the direction from the rotation center.
This artifact makes the treatment of the interaction between the spin and the angular velocity near the rotation center highly nontrivial.  This subtlety might be an explanation of the lattice-QCD results with a very weak dependence on rotation at the rotation center.

In summary, we have demonstrated the advantage of the analytical strong-coupling expansion to get insights into our understanding of the deconfinement phase transition in the presence of imaginary rotation.
Our results are consistent with the model and the perturbative calculations, contrary to the numerical lattice-QCD simulations, although the strong-coupling expansion and the numerical simulation employ the same lattice action.
More analytical investigations along these lines, especially about the proper lattice formulation with the expected periodicity and less lattice artifact around the rotation center would be beneficial for us to resolve the mysterious tension between the numerical lattice-QCD simulation results and others.

\section{Acknowledgement}
The authors thank Maxim~Chernodub for useful discussions and warm hospitality at Institut Denis Poisson, Universit\'{e} de Tours where this work was initiated.  K.F.\ thanks Defu~Hou for drawing our attention to their work based on a similar idea during the completion of this manuscript.
This work was supported by JSPS KAKENHI Grant No.~22H05118 (K.F.), No.~23K22487 (K.F.), No.~24KJ0658 (Y.S.), and JST SPRING Grant No.~JPMJSP2108 (Y.S.). 

\bibliography{SCER}
\bibliographystyle{apsrev4-2}
\end{document}